\documentclass[11pt]{article}

\usepackage[utf8]{inputenc}
\usepackage[T1]{fontenc}
\usepackage{amsmath,amssymb,amsthm,bbm}
\usepackage{graphicx,transparent,eso-pic}
\usepackage[document]{ragged2e}
\usepackage{natbib}
\usepackage{multicol}
\usepackage{url}
\usepackage[margin=2cm]{geometry}

\newcommand{\noun}[1]{\textsc{#1}}

\DeclareMathOperator{\Cov}{Cov}

\newcommand{\Covb}[2]{\ensuremath{\Cov\!\left[#1,#2\right]}}

\newcommand{\rhob}[2]{\ensuremath{\rho\!\left[#1,#2\right]}}

\DeclareMathOperator*{\argmin}{\arg\!\min}

\begin{document}

\title{An Urban Morphogenesis Model Capturing Interactions between Networks and Territories}
\author{\noun{Juste Raimbault}$^{1,2}$\\
$^1$ UPS CNRS 3611 ISC-PIF\\
$^2$ UMR CNRS 8504 G{\'e}ographie-cit{\'e}s
}
\date{}

\maketitle

\justify

\begin{abstract}
Urban systems are composed by complex couplings of several components, and more particularly between the built environment and transportation networks. Their interaction is involved in the emergence of the urban form. We propose in this chapter to introduce an approach to urban morphology grasping both aspects and their interaction. We first define complementary measures, study their empirical values and their spatial correlations on European territorial systems. The behavior of indicators and correlations suggest underlying non-stationary and multi-scalar processes. We then introduce a generative model of urban growth at a mesoscopic scale. Given a fixed exogenous growth rate, population is distributed following a preferential attachment depending on a potential controlled by the local urban form (density, distance to network) and network measures (centralities and generalized accessibilities), and then diffused in space to capture urban sprawl. Network growth is included through a multi-modeling paradigm: implemented heuristics include biological network generation and gravity potential breakdown. The model is calibrated both at the first (measures) and second (correlations) order, the later capturing indirectly relations between networks and territories.
\end{abstract}

\textbf{Keywords : }\textit{Urban Morphology; Road Network Topology; Spatial Correlations; Urban Morphogenesis; Reaction-diffusion; Co-evolution}

\section{Introduction}

\subsection{Urban morphology}

The structure of urban systems determine their functional properties at several scales, and the \emph{urban morphology} both participates in the emergence of urban functions and determines how they evolve and how they are used by the agents \citep{batty1994fractal}. For example, relationships between urban form, mobility practices and sustainability of metropolitan areas can be established \citep{le2010approche}.

Several approaches can be taken to study urban morphology. At the microscopic scale, in operational urbanism for example, urban morphology is defined as ``the characteristics of the material form of cities and fabrics''~\citep{paquot2010abc}. A similar positioning is taken by architecture, considering forms at the scale of buildings \citep{moudon1997urban}. At the mesoscopic scales, \cite{tsai2005quantifying} introduces indicators to quantify urban form, considered as the spatial distribution of population. \cite{le2009quantifier} recalls the necessity of a multi-dimensional measure. It is however possible to obtain a robust description of urban form with a small number of independent indicators by a reduction of the dimension~\citep{Schwarz201029}. It is also possible to use indexes from fractal analysis, such as for example systematically applied by~\cite{2016arXiv160808839C} to classify urban forms. Other more original indexes can be proposed, such as by~\cite{lee2017morphology} which use the variations of trajectories for routes going through a city to establish a classification and show that it is strongly correlated with socio-economic variables. An other entry is through the topology of urban networks, and more particularly road networks, for which \citep{2015arXiv151201268L} gives a broad overview. Finally, geographers also consider at the macroscopic scale the spatial distribution of cities and their populations as the form of the urban system \citep{pumain2011systems}.

The link between urban morphology and the topology of the underlying relational network has been suggested in a theoretical approach by~\cite{badariotti2007conception}. \cite{d2015mathematize} models cities as isobenefit lines for agents, and suggests in opening the link between the distribution of amenities and networks and flows, highlighting the open question of co-evolutive processes between the urban form and spatial networks, in the sense of circular causalities.

There is to the best of our knowledge no contribution in the literature considering the interaction between the built environment and networks as an intrinsic process of the emergence of the urban form. This chapter thus aims at sketching an exploration of this concept at the mesoscopic scale, through empirical data analysis and modeling. More precisely, our contribution is twofold: (i) we introduce indicators quantifying simultaneously the morphology of population spatial distribution and the topology of road networks, and study empirically their values and spatial correlations, including thus the interactions between both in the quantification of the form; (ii) we describe and explore a model of urban morphogenesis, aiming at capturing these interaction processes in the emergence of the form.

The rest of this chapter is organized as follows. We first detail the rationale of our approach to urban form, in particular the scale and objects considered. The second section develops the methodology used to quantify urban form, from the point of view of morphological indicators, network topology indicators, and their correlations. Empirical results for Europe are then described, to then be used to calibrate a model of urban morphogenesis that we detail and explore. We finally discuss potential developments and applications of this work.

\subsection{Urban morphology and interactions between networks and territories}

Through relocation processes, sometimes induced by networks, we can expect the latest to influence the distribution of populations in space \citep{wegener2004land}. Reciprocally, network characteristics can be influenced by this distribution. We propose therefore here to study a coupled approach to the urban form, given by synthetic indicators for these two subsystems, and by correlations between these indicators. At the scale of the system of cities, the spatial nature of the urban system is captured by cities position, associated with aggregated city variables. We propose to work here at the mesoscopic scale, at which the precise spatial distribution of activities is necessary to understand the spatial structure of the territorial system. We will therefore use the term of morphological characteristics for population density and the road network.

The choice of ``relevant'' boundaries for the territory or the city is a relatively open problem which will often depend on the question we are trying to answer \citep{paez2005spatial}. This way, \cite{guerois2002commune} show that the entities obtained are different if we consider an entry by the continuity of the built environment (morphological), by urban functions (employment area for example) or by administrative boundaries. \cite{boeing2017multi} furthermore shows that statistics of network measures significantly change when the scale of study switches from neighborhood to cities and metropolitan areas. Similarly, \cite{cottineau2017diverse} show that scaling exponents also strongly depends on boundaries and thresholds chosen to define the urban area.

To tackle this ontological issue of the object under study, we (i) stay at a high resolution; and (ii) compute field values with a small enough offset to obtain continuous values. At the chosen scale, we can assume that territorial characteristics, for population and network, are locally defined and vary in an approximately continuous way in space. We will compute therefore the indicators on spatial windows of fixed size, taken of the order of magnitude of 100km (in practice 50km, following \cite{2017arXiv170806743R} with furthermore shows a low sensitivity of indicators when comparing with windows of 30 and 100km), but with a small enough offset. Thus, the construction of fields of morphological indicators will allow us to endogenously reconstruct territorial entities through the emergent spatial structure of indicators at larger scales.

\section{Measuring morphology: method}

We detail in this section the indicators used to quantify morphology, both for population distribution (urban morphology) and for network topology.

\subsection{Urban morphology}

Our quantification of the urban form in itself, in the sense of properties of the spatial distribution of populations, is taken from a previous work developed in \cite{2017arXiv170806743R}. We recall here the formal definition of morphological indicators. We work with gridded population data $(P_i)_{1\leq i \leq N^2}$, where $M=N^2$, $d_{ij}$ is the euclidian distance between cells $i,j$, $P=\sum_{i=1}^{M} P_i$ the total population and $\bar{P} = P / M$ the average population per cell. We use the following indicators: (i) rank-size slope $\gamma$, expressing the degree of hierarchy in the distribution; (ii) entropy of the distribution~\citep{le2015forme}, given by 

\begin{equation}
\mathcal{E} = \sum_{i=1}^{M}\frac{P_i}{P}\cdot \ln{\frac{P_i}{P}}
\end{equation}
 
(iii) spatial-autocorrelation given by Moran index~\citep{tsai2005quantifying} defined as

\begin{equation}
I = M \cdot \frac{\sum_{i\neq j} w_{ij} \left(P_i - \bar{P}\right)\cdot\left(P_j - \bar{P}\right)}{\sum_{i\neq j} w_{ij} \sum_{i}{\left( P_i - \bar{P}\right)}^2}
\end{equation}

where the spatial weights are taken as $w_{ij} = 1/d_{ij}$; and (iv) average distance between individuals~\citep{le2009quantifier} given by

\begin{equation}
\bar{d} = \frac{1}{d_M}\cdot \sum_{i<j} \frac{P_i P_j}{P^2} \cdot d_{ij}
\end{equation}

where $d_M$ is a normalisation constant taken as the diagonal of the area on which the indicator is computed in our case.

The first two indexes are not spatial, and are completed by the last two that take space into account. Following \cite{Schwarz201029}, the effective dimension of the urban form justifies the use of all.

\subsection{Network Measures}

We consider network aggregated indicators as a way to characterize transportation network properties on a given territory, the same way morphological indicators yielded information on urban structure. They are assumed thus to capture another dimension of urban form. We propose to compute some simple indicators on same spatial extents as for population density, to be able to explore relations between these static measures.

Static network analysis has been extensively documented in the literature, such as for example through the typology of urban street networks obtained by \cite{louf2014typology} on a cross-sectional study of cities. Similarly, \cite{2017arXiv170902939M} uses techniques from deep learning to establish a typology of urban road networks for a large number of cities across the world. The questions behind such approaches are multiple: they can aim at finding typologies or at characterizing spatial networks, at understanding underlying dynamical processes in order to model morphogenesis, or even at being applied in urban planning such as \emph{Space Syntax} approaches~\citep{hillier1989social}. We aim here at characterizing urban morphological properties.

We introduce indicators to have a broad idea of the form of the network, using a certain number of indicators to capture the maximum of dimensions of properties of networks, more or less linked to their use. These indicators summarize the mesoscopic structure of the network and are computed on topological networks obtained through simplification steps that will be detailed later. If we denote the network with $N=(V,E)$, nodes have spatial positions $\vec{x}(V)$ and populations $p(v)$ obtained through an aggregation of population in the corresponding Voronoï polygon, and edges $E$ have \emph{effective distances} $l(E)$ taking into account impedances and real distances (to include the primary network hierarchy). We then use:

\begin{itemize}
\item Characteristics of the graph, obtained from graph theory, as defined by~\cite{haggett1970network}: number of nodes $\left|V\right|$, number of links $\left|E\right|$, density $d$, average length of links $\bar{d_l}$, average clustering coefficient $\bar{c}$, number of components $c_0$.
\item Measures directly related to distances within the network: diameter $r$, euclidian performance $v_0$ (defined by~\cite{banos2012towards}), average length of shortest paths $\bar{l}$.
\item Centrality measures: these are aggregated at the level of the network by taking their average and their level of hierarchy, computed by an ordinary least squares of a rank-size law, for the following centrality measures:
\begin{itemize}
\item Betweenness centrality~\citep{crucitti2006centrality}, average $\bar{bw}$ and hierarchy $\alpha_{bw}$: given the distribution of centrality on all nodes, we take the slope of a rank-size adjustment and the average of the distribution.
\item Closeness centrality~\citep{crucitti2006centrality}, average $\bar{cl}$ and hierarchy $\alpha_{cl}$.
\item Accessibility~\citep{hansen1959accessibility}, which is in our case computed as a closeness centrality weighted by populations: average $\bar{a}$ and hierarchy $\alpha_{a}$.
\end{itemize}
\end{itemize}

Network performance is close to the rectilinearity measure (\emph{straightness}) proposed by \cite{josselin2016straightness}, which show that it efficiently differentiates rectilinear networks and radio-concentric networks, that are both recurring urban networks. Our indicators are conceived to capture network topology but not the use of the network: developments with suited data could extend these analyses to the functional aspect of networks, such as for example performance measures computed by~\cite{trepanier2009calculation} using massive data for a public transportation network.

\subsection{Correlations}

Local spatial correlations are computed on spatial windows gathering a certain number of observations, and thus of windows on which indicators have been computed. We denote by $l_0$ the resolution of the distribution of indicators. The estimation of correlations in then done on squares of size $\delta\cdot l_0$ (with $\delta$ which can vary typically from 4 to 100). Correlations are estimated using a standard Pearson estimator. Our approach is equivalent to computing geographically weighted correlations \citep{brunsdon2002geographically} with a heaviside window.

$\delta$ gives simultaneously the number of observations used for the local estimation of correlation, and the spatial range of the corresponding window. Its value thus directly influences the confidence of the estimation. We can indeed derive the behavior of the correlation estimator as a function of the size of the sample. Under the assumption of a normal distribution of two random variables $X,Y$, then the Fisher transform of the Pearson estimator $\hat{\rho}$ such that $\hat{\rho} = \tanh (\hat{z})$ has a normal distribution. If $z$ is the transform of the real correlation $\rho$, then a confidence interval for $\rho$ is of size

\begin{equation}
\rho_{+} - \rho_{-} = \tanh (z + k / \sqrt{N}) - \tanh (z - k / \sqrt{N})
\end{equation}

where $k$ is a constant. As $\tanh{z} = \frac{\exp (2z) - 1}{\exp (2z) + 1}$, we can develop this expression and reduce it. We obtain

\begin{equation}
	\rho_{+} - \rho_{-} = 2\cdot \frac{\sinh{(2k/\sqrt{N})}}{\cosh{(2z)} + \cosh{(2k/\sqrt{N})}}
\end{equation}

Using the fact that $\cosh u \sim_0 1 + u^2/2$ and that $\sinh u \sim_0 u$, we indeed obtain that 

\begin{equation}
\rho_{+} - \rho_{-} \sim_{N\gg 0} k' / \sqrt{N}
\label{eq:confidenceinterval}
\end{equation}

This expected asymptotic confidence interval will be of use when studying the behavior of correlations as a function of $\delta$.

\section{Empirical application}

We can now give implementation details and results for the application of this method to local territorial systems of Europe. All source code and results for this section are available on the open repository of the project at \url{https://github.com/JusteRaimbault/CityNetwork/tree/master/Models/StaticCorrelations}.

\subsection{Urban morphology}

As this work extends \cite{2017arXiv170806743R}, the empirical values of morphological indicators for population distribution are taken from it. We recall that the implementation of indicators must be done carefully, since computational complexities can reach $O(N^4)$ for the Moran index for example: we use convolution through Fast Fourier Transform, which is a technique allowing the computation of the Moran index with a complexity in $O(\log^2 N \cdot N^2)$.

Indicators are computed on 50km width square windows with grids of resolution 500m, computed from the Eurostat population dataset \citep{eurostat}. According to \cite{batista2013high} which details its construction, our aggregation should allow us to avoid biases at high resolutions. The spatial distribution of indicators unveils typical local and regional regimes, which \cite{2017arXiv170806743R} synthesizes using unsupervised learning. They typically contain a metropolitan regime, a medium-sized city regime, a rural regime (split in two between North and South), a mountainous regime.

\subsection{Network Topology}

\subsubsection{Data preprocessing}

The implementation of network indicators require a preprocessing from raw data, that we detail first. We assume to work only with the road network. Indeed, data for the current road network is openly available through the OpenStreetMap (OSM) project~\citep{openstreetmap}. Its quality was investigated for different countries such as England~\citep{haklay2010good} and France~\citep{girres2010quality}. It was found to be of a quality equivalent to official surveys for the primary road network. We will however simplify the network at a sufficient level of aggregation to ensure the robustness of results.

The network constituted by primary road segments is aggregated at the fixed granularity of the density grid to create a graph. It is then simplified to keep only the topological structure of the network, normalized indicators being relatively robust to this operation. This step is necessary for a simple computation of indicators and a thematic consistence with the density layer.

Recent tools such as the one proposed by \cite{boeing2017osmnx} provide algorithms to operate an extraction of network topology. The algorithm we use is very similar but is necessary as our approach necessitates special tuning for the following points: (i) aggregation of data at the level of raster cells, which resolution can be variable; and (ii) construction of networks on significant areas (continental scale), made possible in our case by a parallelization of the computation.

We keep only the nodes with a degree strictly greater or smaller than two, and corresponding links, by taking care to aggregate the real geographical distance when constructing the corresponding topological link. Given the order of magnitude of data size (for Europe, the initial database has $\simeq 44.7\cdot 10^6$ links, and the final simplified database $\simeq 20.4\cdot 10^6$), a specific parallel algorithm is used, with a \emph{split-merge} structure. It separates the space into areas that can be independently processed and then merged. We detail it in the following.

\subsubsection{Network Simplification Algorithm}

We detail here the road network simplification algorithm from OpenStreetMap data. The general workflow is the following: (i) data import by selection and spatial aggregation at the raster resolution; (ii) simplification to keep only the topological network, processed in parallel through \emph{split/merge}.

OSM data are imported into a \texttt{pgsql} database (\texttt{Postgis} extension for the management of geometries and to have spatial indexes). The import is done using the software \texttt{osmosis}~\cite{osmosis}, from an image in compressed \texttt{pbf} format of the OpenStreetMap database (the dump was retrieved from \url{http://download.geofabrik.de}, in July 2016). We filter at this stage the links (\texttt{ways}) which posses the tag \texttt{highway}, and keep the corresponding nodes.

The network is first aggregated at a 100m granularity in order to be consistently used with population grids. It furthermore allows to be robust to local coding imperfections or to very local missing data. For this step, roads are filtered on a relevant subset of tags, that we take within \texttt{motorway}, \texttt{trunk}, \texttt{primary}, \texttt{secondary}, \texttt{tertiary}, \texttt{unclassified}, \texttt{residential}. For the set of segments of corresponding lines, a link is created between the origin and the destination cell, with a real length computed between the center of cells and a speed taken as the speed of the line if it is available.

The simplification is then operated the following way:
\begin{enumerate}
\item The whole geographical coverage is cut into areas on which computations will be partly done through parallel computation (\emph{split} paradigm). Areas have a fixed size in number of cells of the base raster (200 cells).
\item On each sub-area, a simplification algorithm is applied the following way: as long as there still are vertices of degree 2, successive sequences of such vertices are determined, and corresponding links are replaced by a unique link with real length and speed computed by cumulation on the deleted links.
\item As the simplification algorithm keeps the links having an intersection with the border of areas, a fusion followed by a simplification of resulting graphs is necessary. To keep a reasonable computational cost, the size of merged areas has to stay low: we take merge areas composed by two contiguous areas. A paving by four sequences of independent merging allows then to cover the full set of joints between areas, these sequences being executed sequentially.
\end{enumerate}

We have then at our disposition a topological graph given by the links between cells of the base raster, having distance and speed attributes corresponding to the underlying real links. This topological graph for Europe has been made available as an open database on the dataverse repository at \url{https://doi.org/10.7910/DVN/RKDZMV}.

\subsubsection{Empirical values of network indicators}

\begin{figure}
\includegraphics[width=\linewidth]{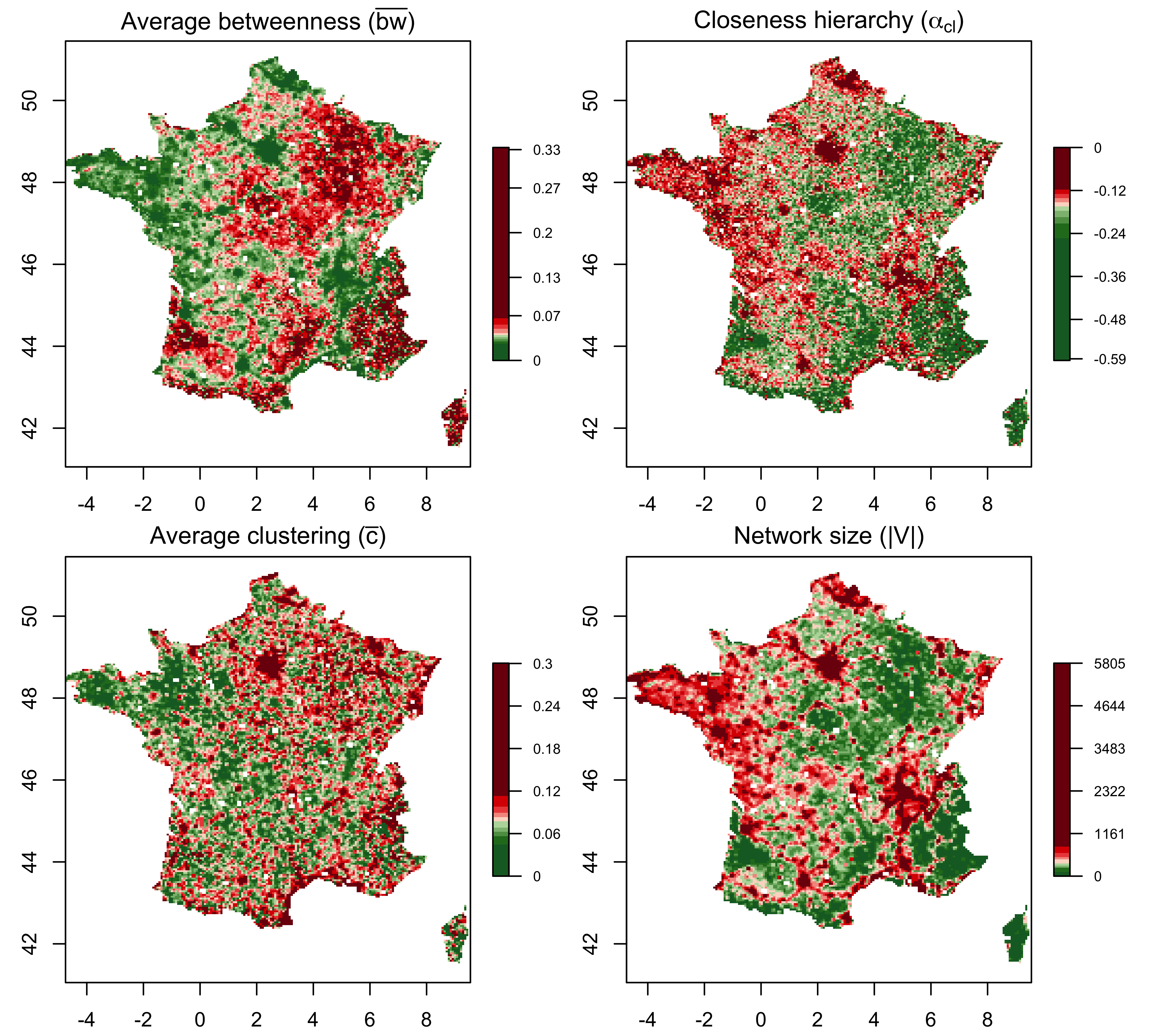}
\caption{\textbf{Spatial distribution of network indicators.} We show indicators for France, in correspondance with morphological indicators described previously. We give here the average betweenness centrality $\bar{bw}$, the hierarchy of closeness centrality $\alpha_cl$, the average clustering coefficient $\bar{c}$ and the number of nodes $\left|V\right|$.\label{fig:staticcorrs:network}}
\end{figure}

Network indicators have been computed on the same areas as urban form indicators, in order to put them in direct correspondance and later compute the correlations. We show in Fig.~\ref{fig:staticcorrs:network} a sample for France.

The spatial behavior of indicators unveils local regimes as for the urban form (urban, rural, metropolitan), but also strong regional regimes. They can be due to the different agricultural practices depending on the region for the rural for example, implying a different partition of parcels and also a particular organization of their serving. For network size, Brittany is a clear outlier and rejoins urban regions, witnessing very fragmented parcels (and a fortiori also of a land property fragmentation in the simplifying assumption of corresponding parcels and properties). This is partly correlated to a low hierarchy of accessibility. The South and the East of the extended \emph{Bassin Parisien} are distinguishable by a strong average betweenness centrality, in accordance with a strong hierarchy of the network.

The same way as for urban form, this spatial variability suggests the search of variables regimes of interactions between indicators, as we will do for later through their correlations.

\subsection{Effective static correlations and non-stationarity}

\subsubsection{Spatial correlations}

\begin{figure}
\includegraphics[width=\linewidth]{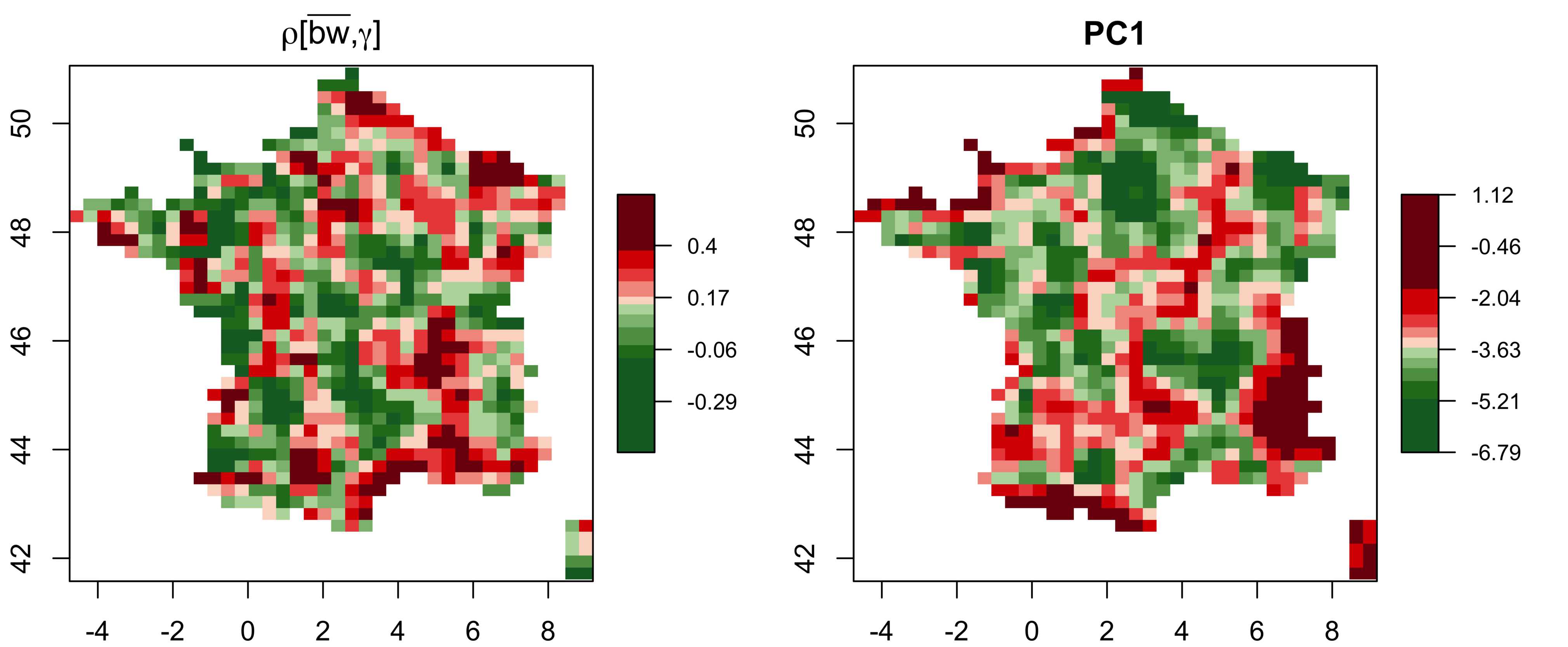}
\caption{\textbf{Examples of spatial correlations.} For France, the maps give $\rho\left[\bar{bw},\gamma\right]$, correlation between the average betweenness centrality and the hierarchy of population (\textit{Left}) and the first component of the reduced matrix (\textit{Right}).\label{fig:staticcorrs:mapscorrs}}
\end{figure}

\begin{figure}
	\includegraphics[width=\linewidth]{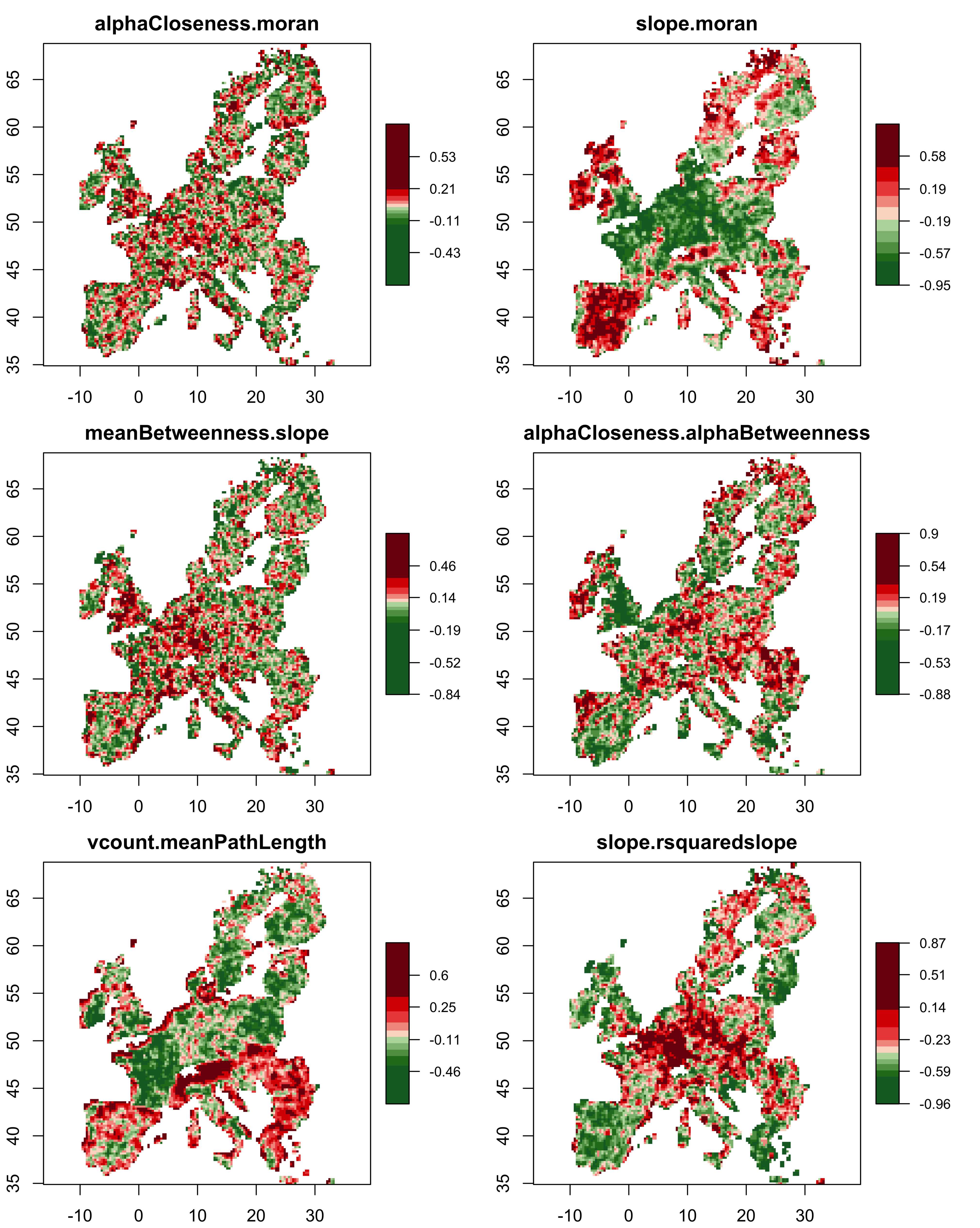}
\caption{\textbf{Spatial correlations for Europe.} The estimation is done here with $\delta = 12$.\label{fig:app:staticcorrelations:europe-correlations}}
\end{figure}

We show in Fig.~\ref{fig:staticcorrs:mapscorrs} examples of correlations estimated with $\delta = 12$ in the case of France. With 20 indicators, the correlation matrix is significantly large in size, but the effective dimension (the number of components required to reach the majority of variance) is reduced:  principal components analysis shows that 10 components already capture 62\% of the variance, and the first component already captures 17\%, what is considerable in a space where the dimension is 190.

The Fig.~\ref{fig:app:staticcorrelations:europe-correlations} gives the spatial distribution for all Europe, for a sample of correlations between indicators: $\rhob{\alpha_{cl}}{I}$, $\rhob{\gamma}{\alpha}$, $\rhob{\bar{bw}}{\gamma}$, $\rhob{\alpha_{bw}}{\alpha_{cl}}$, $\rhob{\left|V\right|}{\bar{l}}$, $\rhob{\gamma}{r_{\gamma}}$ (with $r_{\gamma}$ adjustment coefficient for $\gamma$). We see interesting structures emerging, such as the hierarchy and its adjustment which exhibit an area of strong correlation in the center of Europe and negative correlation areas, or the number of nodes and the path length which correlate in mountains and along the coasts (what is expected since roads do several detours in such topographies) and have a negative correlation otherwise.

It is possible to examine within the correlation matrix the bloc for urban form, for the network, or for crossed correlations, which directly express a link between properties of the urban form and of the network. For example, a certain correspondence between average betweenness centrality and morphological hierarchy that we obtain allows to understand the process corresponding to the correspondance of hierarchies: a hierarchical population can induce a hierarchical network or the opposite direction, but it can also induce a distributed network or such a network create a population hierarchy - this must be well understood in terms of correspondence and not causality, but this correspondance informs on different urban regimes. Metropolitan areas seem to exhibit a positive correlation for these two indicators, as shows the Fig.~\ref{fig:staticcorrs:mapscorrs}, and rural spaces a negative correlation.

In order to give a picture of global relations between indicators, we can refer to the matrix obtained for $\delta = \infty$: for example, a strong population hierarchy is linked to a high and hierarchical betweenness centrality, but is negatively correlated to the number of edges (a diffuse population requires a more spread network to serve all the population). However, it is not possible this way to systematically link indicators, since they especially strongly vary in space.

Furthermore, to give an idea of the robustness of the estimation, we investigate for the correlations estimated on the full dataset (corresponding to $\delta = \infty$) the relative size of confidence intervals at the 95\% level (Fisher method) given by $\frac{\left|\rho_+ - \rho_-\right|}{\left|\rho\right|}$, for correlations such that $\left|\rho\right|>0.05$. The median of this rate is at $0.04$, the ninth decile at $0.12$ and the maximum at $0.19$, what means that the estimation is always relatively good compared to the value of correlations.

This suggests a very high variety of interaction regimes. The spatial variation of the first component of the reduced matrix confirms it, what clearly reveals the spatial non-stationarity of interaction processes between forms, since the first and second moments vary in space. The statistical significance of stationarity can be verified in different ways, and there does not exist to the best of our knowledge a generic test for spatial non-stationarity. \cite{zhang2014test} develop for example a test for rectangular regions of any dimension, but in the specific case of \emph{point processes}. We use here the method of \cite{leung2000statistical} which consists in estimating through bootstrap the robustness of Geographically Weighted Regression models. These will be developed below, but we obtain for all tested models a significant non-stationarity without doubt ($p<10^{-3}$).

\subsubsection{Variations of the estimated correlations}

\begin{figure}[h!]
\includegraphics[width=\linewidth,height=0.85\textheight]{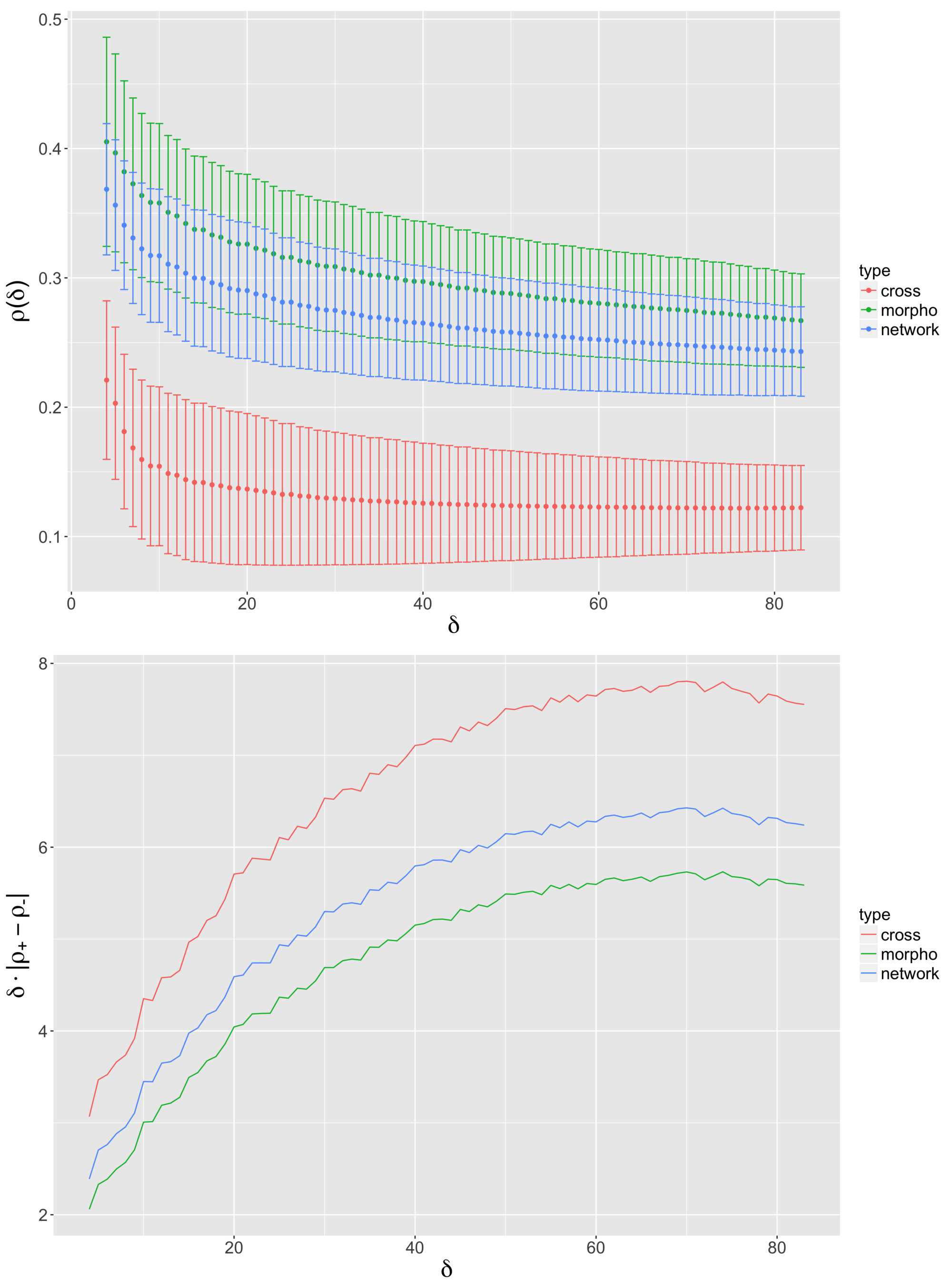}
\caption{\textbf{Variation of correlations with scale, for correlations computed on Europe.} (\textit{Top}) Average absolute correlations and their standard deviations, for the different blocs, as a function of $\delta$; (\textit{Bottom}) Normalized size of the confidence interval $\delta\cdot \left|\rho_{+} - \rho_{-}\right|$ (confidence interval $\left[\rho_{-} , \rho_{+}\right]$ estimated by the Fisher method) as a function of $\delta$.\label{fig:staticcorrs:corrsdistrib}}
\end{figure}

We show in Fig.~\ref{fig:staticcorrs:corrsdistrib} the variation of the estimation of correlation as a function of window size. More precisely, we observe a strong variation of correlations as a function of $\delta$, what is reflected in the average value of the matrix given here (which extends for example from $\rho(4)=0.22$ to $\rho(80)=0.12$ for average absolute cross-correlations). An increase of $\delta$ leads for all measures a shift towards positive values, but also a narrowing of the distribution, these two effects resulting in a decrease of average absolute correlations, which approximatively stabilize for large values of $\delta$.

Such a variation could be a clue of a multi-scalar behavior: a change in window size should not influence the estimation if a single process would be implied, it should only change the robustness of the estimation. Let sketch in a simplified case how this link can be done. To simplify, we consider processes with two characteristic scales which linearly superpose, i.e. which can be written as

\begin{equation}
X_i = X_i^{(0)} + \tilde{X}_i
\end{equation}

with $X_i^{(0)}$ trend at the small scales with a characteristic evolution distance $d_0$, and $\tilde{X}_i$ signal evolving at a characteristic distance $d \ll d_0$. We can then compute the decomposition of the correlation between two processes. We assume that $\Covb{X_i^{(0)}}{\tilde{X}_j} = 0$ for all $i,j$, and denoting $\varepsilon_i = \frac{\sigma\left[X_i^{(0)}\right]}{\sigma\left[\tilde{X}_i\right]}$ the rate between standard deviations of trend and signal. Developing the correlation $\rho\left[X_1,X_2\right]$ by bilinearity and developing the resulting expression at the first order under the assumption that $\varepsilon_i \ll 1$ gives the following approximation

\begin{equation}
	\rho\left[X_1,X_2\right] = \left( \varepsilon_1 \varepsilon_2\rho\left[X_1^{(0)},X_2^{(0)}\right] + \rho\left[\tilde{X}_1,\tilde{X}_2\right]\right)\cdot\left(1 - \frac{1}{2}(\varepsilon_1^2 + \varepsilon_2^2)\right)
	\label{eq:correlations}
\end{equation}

The overall correlation is thus corrected by both an attenuation and an interference factor.

Let apply this result to our problematic. We observe the following reasonable assumptions: (i) for areas with a size smaller than the stationarity scale of correlations (which we have shown empirically to exist at least for some indicators), estimating the covariance of noise should be equivalent on any smaller window in terms of estimator values; (ii) trends, if they exist, have a very low variance, i.e. $\varepsilon_i \ll 1$; (iii) trends are uncorrelated. Combining these three hypotheses with Eq.~\ref{eq:correlations} implies that $\rho (\delta)$ should decrease with $\delta$ if they are verified.

This confirms in this ideal case the observed empirical variation in Fig.~\ref{fig:staticcorrs:corrsdistrib}. A formal demonstration of this hypothesis for more general type of processes and less restricting assumptions remains however out of the scope of this work.

An other signature of a possible multi-scalarity is shown by the variation of the normalized size of the confidence interval for correlations, which in theory under an assumption of normality should lead $\delta\cdot \left|\rho_+ - \rho -\right|$ to remain constant, since boundaries vary asymptotically as $1/\sqrt{N}\sim 1/\sqrt{\delta^2}$ as obtained in Eq.~\ref{eq:confidenceinterval}, follows the direction of this hypothesis of processes superposed at different scales as proposed previously.

Thus, processes are non-stationary, and this stylized insight suggests that they are the product of the superposition of processes at different scales. We however recall that the notion of multi-scalar process is otherwise very broad, and can manifest itself in scaling laws for example~\citep{west2017scale}. An approach closer to the one we took is given by~\cite{Chodrow31102017} which measures intrinsic scales to segregation phenomenons by using measures from Information Theory.

\subsubsection{Typical scales}

\begin{table}[h!]
\caption{\textbf{Interrelations between network indicators and morphological indicators.} Each relation is adjusted by a Geographically Weighted Regression, for the optimal range adjusted by AICc.\label{tab:staticcorrelations:gwr}}
\begin{center}
\begin{tabular}{|l|l|l|l|}
\hline
Indicator & Model & Range (km) & Adjustment ($R^2$) \\ \hline
Average distance $\bar{d}$ & $\bar{d} \sim v_0$ & 11.6 & 0.31 \\
Entropy $\mathcal{E}$  & $\mathcal{E} \sim v_0$ &  8.8  &0.75 \\
Moran $I$ & $I \sim v_0$ & 8.8 & 0.49 \\
Hierarchy $\gamma$ & $\gamma \sim v_0$ & 8.8  & 0.68 \\\hline
Average betweenness $\bar{bw}$ & $\bar{bw} \sim I$ & 12.3 & 0.58 \\
Average closeness $\bar{cl}$ & $\bar{cl}\sim I$ & 13.9 & 0.26 \\
Performance $v_0$ & $v_0 \sim \mathcal{E}$ & 8.6  & 0.86 \\
Number of nodes $\left|V\right|$ & $\left|V\right| \sim \mathcal{E}$ & 8.6  & 0.88 \\\hline
\end{tabular}
\end{center}
\end{table}

We also propose to explore the possible property of multi-scalar processes by the extraction of endogenous scales in the data. A Geographically Weighted Principal Component Analysis (GWRPCA)~\cite{harris2011geographically} as exploratory analysis suggests weights and importances that vary in space, what is in consistence with the non-stationarity of correlation structures obtained above. There is no reason a priori that the scales of variation of the different indicators are strictly the same. We propose thus to extract typical scales for crossed relations between the urban form and network topology.

We implement therefore the following method: we consider a typical sample of indicators (four for each aspect, see the list in Table~\ref{tab:staticcorrelations:gwr}), and for each indicator we formulate all the possible linear models as a function of opposite indicators (network for a morphological indicator, morphological for a network indicator), aiming at directly capturing the interaction without controlling on the type of form or of network. These models are then adjusted by a Geographically Weighted Regression (GWR) with an optimal range determined by a corrected information criteria (AICc), using the R package GWModel~\citep{gollini2013gwmodel}. For each indicator, we keep the model with the best value of the information criteria. We adjust the models on data for France, with a \emph{bisquare} kernel and an adaptative bandwidth in number of neighbors.

Results are presented in Table~\ref{tab:staticcorrelations:gwr}. It is first interesting to note that all models have only one variable, suggesting relatively direct correspondances between topology and morphology. All morphological indicators are explained by network performance, i.e. the quantity of detours it includes. On the contrary, network topology is explained by Moran index for centralities, and by entropy for performance and the number of vertices. There is thus a dissymmetry in relations, the network being conditioned in a more complex way to the morphology than the morphology to the network. The adjustments are rather good ($R^2 > 0.5$) for most indicators, and \emph{p-values} obtained for all models (for the constant and the coefficient) are lower than $10^{-3}$. Concerning the scales corresponding to the optimal model, they are very localized, of the order of magnitude of ten kilometers, i.e a larger variation than the one obtained the correlations. This analysis confirms thus statistically on the one hand the non-stationarity, and on the other hand give a complementary point of view on the question of endogenous scales.

\section{Urban morphogenesis model}

After having characterized empirically urban morphology from the point of view of interactions between the built environment and networks, we propose to gain indirect knowledge on processes involved in the emergence of the form through modeling, by introducing a model of urban morphogenesis which aims at capturing empirical results obtained above.

\subsection{Model rationale}

Urban settlements and transportation networks have been shown to be co-evolving, in the different thematic, empirical  and modeling studies of territorial systems developed up to here. As we saw, modeling approaches of such dynamical interactions between networks and territories are poorly developed. We propose in this section to realize a first entry at an intermediate scale, focusing on morphological and functional properties of the territorial system in a stylized way. We introduce a stochastic dynamical model of urban morphogenesis which couples the evolution of population density within grid cells with a growing road network.

\begin{figure}
	\includegraphics[width=\linewidth]{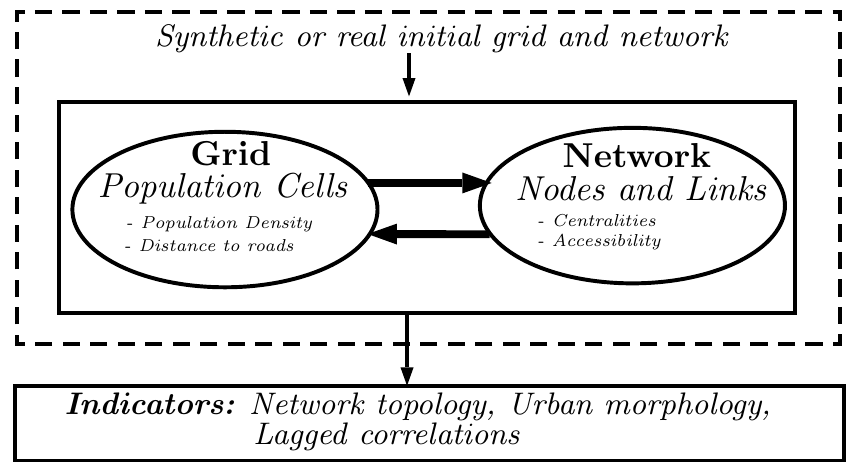}
	\caption{\textbf{Structure of the co-evolution model at the mesoscopic scale.}\label{fig:mesocoevolmodel:workflow}}
\end{figure}

The general principles of the model are the following. With an overall fixed growth rate, new population aggregate preferentially to a local potential, for which parameters control the dependence to various explicative variables. These are in particular local density, distance to the network, centrality measures within the network and generalized accessibility. \cite{doi:10.1080/13658816.2014.893347} shows in the case of Stockholm the very strong correlation between centrality measures in the network and the type of land-use, what confirms the importance to consider centralities as explicative variables for the model at this scale. We generalize thus the morphogenesis model studied in~\cite{2017arXiv170806743R}, with aggregation mechanisms similar to the ones used by~\cite{raimbault2014hybrid}. A continuous diffusion of population completes the aggregation to translate repulsion processes generally due to congestion. Because of the different time scales of evolution for the urban environment and for networks, the network grows at fixed time steps: a first fixed rule ensures connectivity of newly populated patches to the existing network. Different network generation heuristics are then included in the model, which are expected to be complementary. The Fig.~\ref{fig:mesocoevolmodel:workflow} summarizes the general structure of the morphogenesis model.

\subsection{Model description}

The model is based on a squared population grid of size $N$, whose cells are defined by populations $(P_i)$. A road network is included as a vectorial layer on top of the grid, similarly to \cite{raimbault2014hybrid}. We assume at the initial state a given population distribution and a network.

The evolution of densities is based on a utility function, influenced by local characteristics of the urban form and function, that we call \emph{explicative variables}. Let $x_k(i)$ be a local explicative variable for cell $i$, which will be among the following variables: population $P_i$; proximity to roads, taken as $\exp (-d / d_n)$ where $d$ is the distance by projection on the closest road, and $d_n=10$ is fixed; betweenness centrality; closeness centrality; accessibility.

For the last three variables, they are defined as previously for network nodes, and then associated to cells by taking the value of the closest node, weighted by a decreasing function of the distance to it, i.e. of the form $x_k = x^{(n)}_k (\argmin_j d(i,j)) \cdot \exp \left( -  \min_j d(i,j) / d_0 \right)$, with $x^{(n)}_k$ the corresponding variable for nodes, the index $j$ being taken on all nodes, and the decay parameter $d_0$ is in our case fixed at $d_0=1$ to keep the property that network variables are essentially significant at close distances from the network. We consider then normalized explicative variables defined by $\tilde{x}_k(i) = x_k(i) - \min_j x_k(j) / (\max_j x_k(j) - \min_j x_k(j))$.

The utility of a cell is then given by a linear aggregation following

\begin{equation}
U_i = \sum_k w_k \cdot \tilde{x}_k(i)
\end{equation}

where $\tilde{x}_k$ are the normalized local explicative variables, and $w_k$ are weight parameters, which allow to weight between the different influences. Alternatives to this simple utility function include for example Cobb-Douglas functions, what are equivalent to a linear aggregation on the logarithms of variables.

A time step of model evolution includes then the following stages.
\begin{enumerate}
	\item Evolution of the population following rules similar to \cite{2017arXiv170806743R}. Given an exogenous growth rate $N_G$, individuals are added independently following an aggregation done with a probability $U_i^\alpha/\sum_k U_k^\alpha$, followed by a diffusion of strength $\beta$ to neighbor cells, done $n_d$ times.
	\item Network growth is done through multi-modeling as proposed by \cite{raimbault2017modeling} and explored by \cite{raimbault2018multi}, knowing that this takes place is the time step is a multiple of a parameter $t_N$, which allows to integrate a differential between temporal scales for population growth and for network growth. The different heuristic used for adding links are: (i) nothing (baseline); (ii) random links; (iii) deterministic potential breakdown; (iv) random potential breakdown \citep{schmitt2014modelisation}; (v) cost-benefits \citep{louf2013emergence}; (vi) biological network growth \citep{tero2010rules}.
\end{enumerate}

The fact that the aggregation of population follows a power law of the utility provides a flexibility in the underlying optimization problem by agents, since as \cite{josselin2013revisiting} recall, the use of different norms in spatial optimal location problems corresponds to different logics of optimization.

The parameters of the model that we will make vary are then: (i) aggregation-diffusion parameters $\alpha,\beta,N_g,n_d$; (ii) the four weight parameters $w_k$ for the explicative variables, which vary in $[0;1]$; network growth parameters for the different heuristics (see \cite{raimbault2018multi}). Output model indicators are the urban morphology indicators, topological network indicators, and lagged correlations between the different explicative variables.

\subsection{Simulation results}

The model is implemented in NetLogo, given the heterogeneity of aspects that have to be taken into account, and this language being particularly suitable to couple a grid of cells with a network. Urban morphology indicators are computed thanks to a NetLogo extension specifically developed. Source code and results for the modeling part are available at \url{https://github.com/JusteRaimbault/CityNetwork/tree/master/Models/MesoCoevol}.

We propose to focus on the ability of the model to capture relations between networks and territories, and more particularly the co-evolution. Therefore, we will try to establish if (i) the model is able to reproduce, beyond the form indicators, the static correlation matrices computed previously; and (ii) the model produces a variety of dynamical relations in the sense of causality regimes developed by~\cite{raimbault2017identification}.

The model is initialized on fully synthetic configurations, with a grid of size $50$. Configurations are generated through an exponential mixture in a way similar to \cite{anas1998urban}: $N_c = 8$ centers are randomly located, to which a population is attributed following a scaling law $P_i = P_0\cdot (i+1)^{-\alpha_S}$ with $\alpha_S = 0.8$ and $P_0 = 200$. The population of each center is distributed to all cells with an exponential kernel of shape $d(r) = P_{max}\exp\left( - r / r_0\right)$ where the parameter $r_0$ is determined to fix the population at $P_i$, with $P_{max} = 20$ (density at the center). We have indeed $P_i = \iint d(r) = \int_{\theta=0}^{2\pi} \int_{r=0}^{\infty} d(r) rdrd\theta = 2 \pi P_{max} \int_r r\cdot \exp\left( - r / r_0\right) = 2 \pi P_{max} r_0^2$, and therefore $r_0 = \sqrt{\frac{P_i}{2\pi P_{max}}}$. An initial network skeleton is generated by sequential closest components connection.

We explore a Latin Hypercube Sampling of the parameter space, with 10 repetitions for around 7000 parameter points, corresponding to a total of around 70000 model repetitions, realized on a computation grid by using the OpenMole model exploration software \citep{reuillon2013openmole}. The simulation results are available at \url{http://dx.doi.org/10.7910/DVN/OBQ4CS}.

\subsubsection{Static and dynamical calibration}

\begin{figure}
	\includegraphics[width=\linewidth]{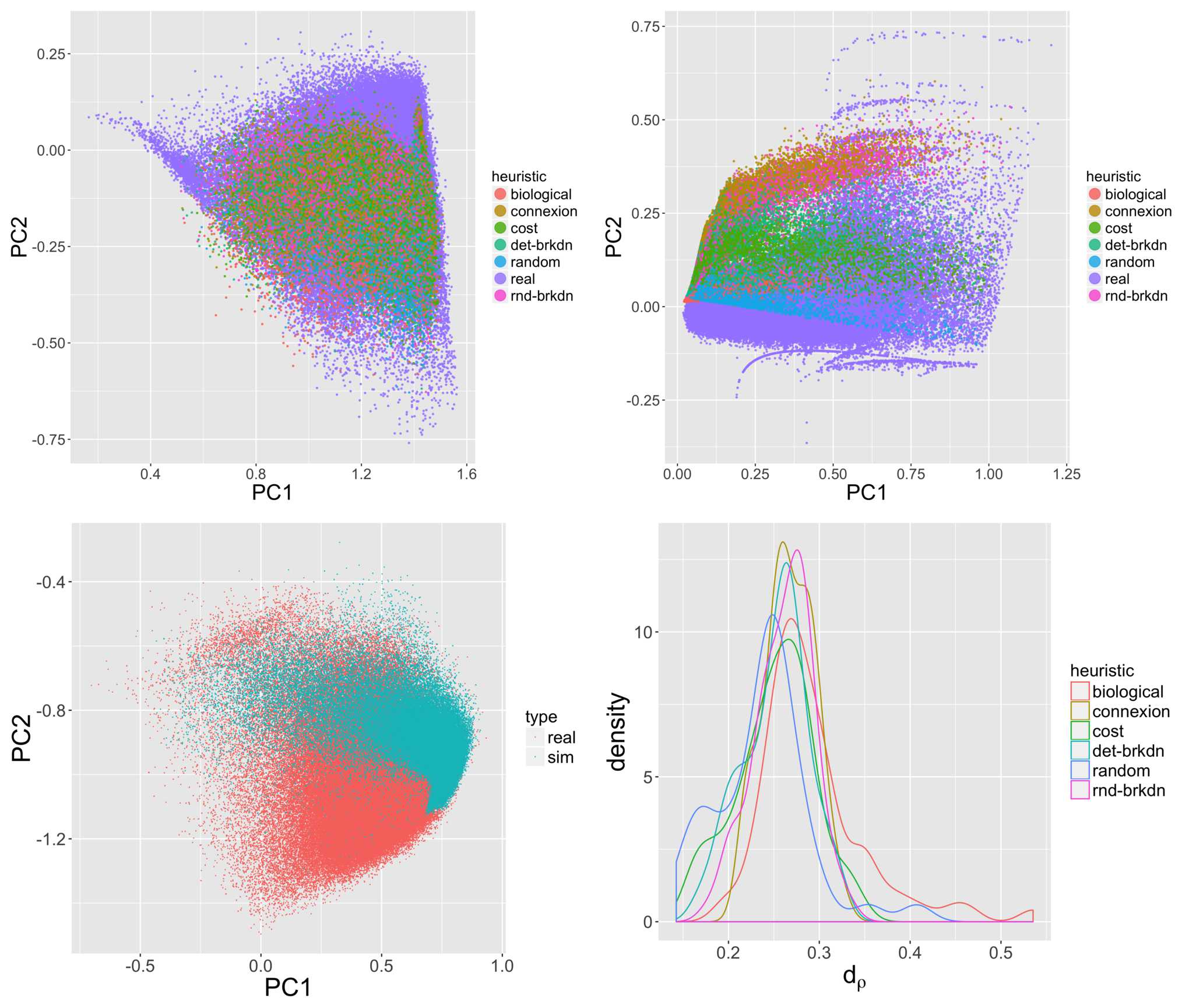}
	\caption{\textbf{Calibration of the morphogenesis model at the first and second order.} (\textit{Top Left}) Simulated and observed point clouds in a principal plan for urban morphology indicators. (\textit{Top Right}) Simulated and observed could points in a principal plan for network indicators. (\textit{Bottom Left}) Simulated and observed point clouds in a principal plan for all indicators. (\textit{Bottom Right}) Distributions of distances on correlations $d_{\rho}$, for the different heuristics.\label{fig:mesocoevolmodel:calibration}}
\end{figure}

The model is calibrated at the first order, on indicators for the urban form and network measures, and at the second order on correlations between these. We introduce an \emph{ad hoc} calibration procedure in order to take into account the first two moments, that we detail below. More elaborated procedures are used for example in economics, such as \cite{watson1993measures} which uses the noise of the difference between two variables to obtain the same covariance structure for the two corresponding models, or in finance, such as \cite{frey2001copulas} which define a notion on equivalence between latent variables models which incorporates the equality of the interdependence structure between variables. We avoid here to add supplementary models, and consider simply a distance on correlation matrices. The procedure is the following.
\begin{itemize}
	\item Simulated points are the ones obtained through the sampling, with average values on repetitions.
	\item In order to be able to estimate correlation matrices between indicators for simulated data, we make the assumption that second moments are continuous as a function of model parameters, and split for each heuristic the parameter space into areas to group parameter points (each parameter being binned into $15 / k$ equal segments, where $k$ is the number of parameters: we empirically observed that this allowed to always have a minimal number of points in each area), what allows to estimate for each group indicators and the correlation matrix.
	\item For each estimation done this way, that we write $\bar{S}$ (indicators) and $\rho [S]$ (correlations), we can then compute the distance to real points on indicators $d_I (R_j) = d(\bar{S},R_j)$ and on correlation matrices $d_{\rho} (R_j) = d(\rho [S],\rho[R_j])$ where $R_j$ are the real points with their corresponding correlations, and $d$ an euclidian distance normalized by the number of components.
	\item We consider then the aggregated distance defined as $d_A^2 (R_j) = d_I^2 (R_j) + d_{\rho}^2 (R_j)$. Indeed, the shape of Pareto fronts for the two distances considered suggests the relevance of this aggregation. The real point closest to a simulated point is then the one in the sense of this distance.
\end{itemize}

The Fig.~\ref{fig:mesocoevolmodel:calibration} summarizes calibration results. Morphological indicators are easier to approach than network indicators, for which a part of the simulated clouds does not superpose with observed points. We find again a certain complementarity between network heuristics. When considering the full set of indicators, few simulated points are situated far from the observed points, but a significant proportion of these is beyond the reach of simulation. Thus, the simultaneous capture of morphology and topology is obtained at the price of less precision.

We however obtain a good reproduction of correlation matrices as shown in Fig.~\ref{fig:mesocoevolmodel:calibration} (histogram for $d_{\rho}$, bottom right). The worse heuristic for correlations is the biological one in terms of maximum, whereas the random produces rather good results: this could be due for example to the reproduction of very low correlations, which accompany a structure effect due to the initial addition of nodes which imposes already a certain correlation. On the contrary, the biological heuristic introduces supplementary processes which can possibly be beneficial to the network in terms of independence (or following the opposed viewpoint be detrimental in terms of correlations). In any case, this application shows that our model is able to resemble real configurations both for indicators and their correlations.

\subsubsection{Causality regimes}

\begin{figure}
	\includegraphics[width=\linewidth,height=0.93\textheight]{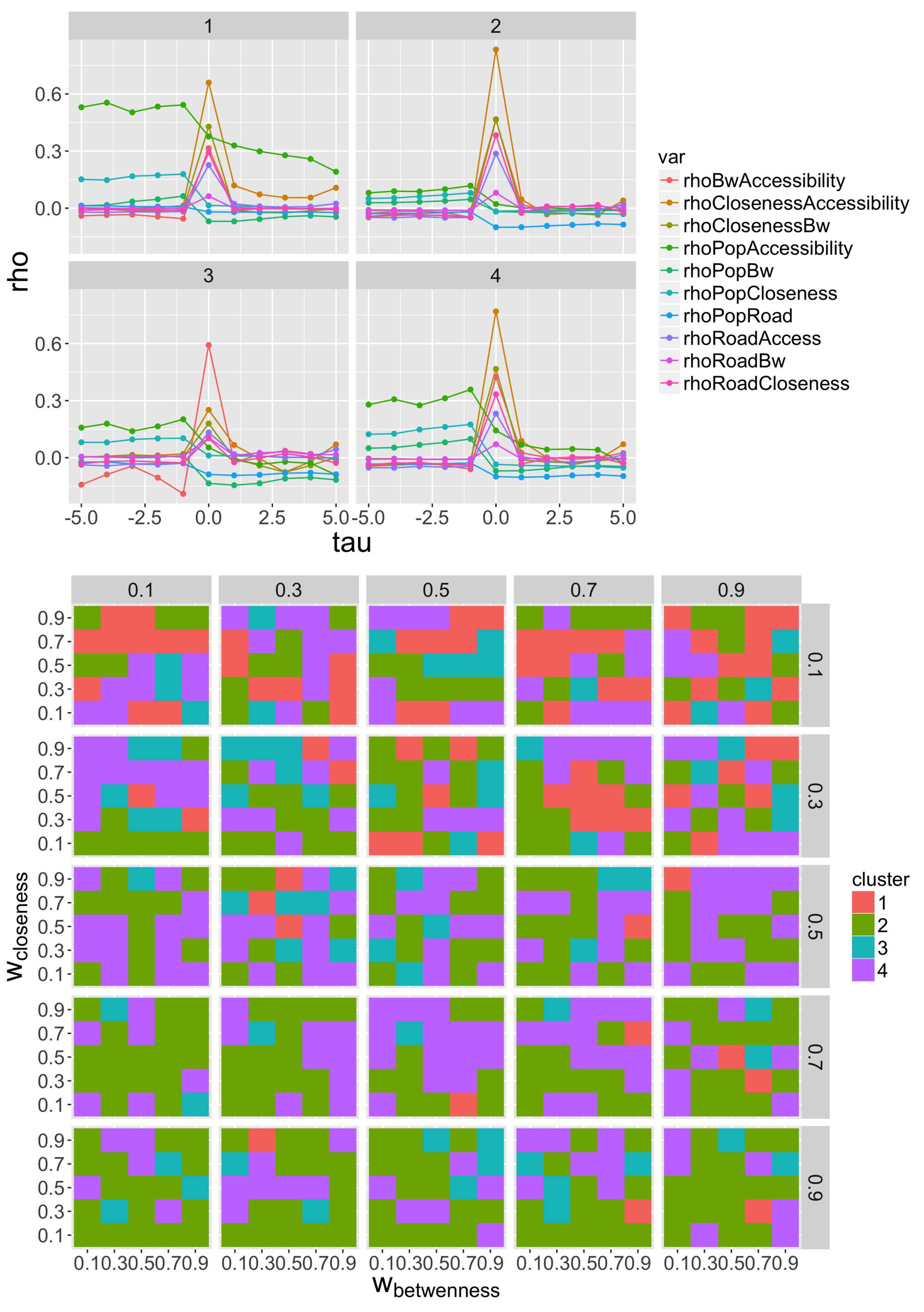}
	\caption{\textbf{Causality regimes for the co-evolution model.} (\textit{Top}) Trajectories of classes centers in terms of $\rho[\tau]$ between the different explicative variables. (\textit{Bottom}) Phase diagram of regimes in the parameter space for $w_k$, represented here as the variation of diagrams for $(w_{bw},w_{cl})$, along the variations of $w_{road}$ (in rows) and of $w_{pop}$ (in columns).\label{fig:mesocoevolmodel:causality}}
\end{figure}

We furthermore study dynamical lagged correlations between the variations of the different explicative variables for cells (population, distance to the network, closeness centrality, betweenness centrality, accessibility). We apply the method of causality regimes introduced in~\cite{raimbault2017identification}. The Fig.~\ref{fig:mesocoevolmodel:causality} summarizes the results obtained with the application of this method on simulation results of the co-evolution model. The number of classes inducing a transition is smaller than for the RDB model, translating a smaller degree of freedom, and we fix in that case $k=4$. Centroid profiles allow to understand to ability of the model to more or less capture a co-evolution.

The regimes obtained appear to be less diverse than the ones obtained by \cite{raimbault2017identification} or for a macroscopic model of co-evolution by~\cite{2018arXiv180409430R}. Some variables have naturally a strong simultaneous correlation, spurious from their definitions, such as closeness centrality and accessibility, or the distance to the road and the closeness centrality. For all regimes, population significantly determines the accessibility. The regime 1 corresponds to a full determination of the network by the population. The second is partly circular, through the effect of roads on populations. The regime 3 is more interesting, since closeness centrality negatively causes the accessibility: this means that in this configuration, the coupled evolution of the network and the population follow the direction of a diminution of congestion. Furthermore, as population causes the closeness centrality, there is also circularity and thus co-evolution in that case. When we locate it in the phase diagram, this regime is rather sparse and rare, contrary for example to the regime 1 which occupies a large portion of space for a low importance of the road ($w_{road} \leq 0.3$). This confirms that the co-evolution produced by the model is localized and not a characteristic always verified, but that it is however able to generate some in particular regimes.

\section{Discussion}

\subsection{Quantifying urban form}

We have first shown empirically the non-stationarity of interactions between the morphology of the distribution of populations and the topology of the road network. Various developments of this analysis are possible.

Population density grids exist for all regions of the world, such as for example the ones provided by~\cite{10.1371/journal.pone.0107042}. The analysis may be repeated on other regions of the world, to compare the correlation regimes and test if urban system properties stay the same, keeping in mind the difficulties linked to the differences in data quality.

The research of local scales, i.e. with an adaptive estimation window in terms of size and shape for correlations, would allow to better understand the way processes locally influence their neighborhood. The validation criteria for window size would still be to determine: it can be as above an optimal range for explicative models that are locally adjusted.

The question of ergodicity, crucial to urban systems \citep{pumain2012urban}, should also be explored from a dynamical point of view, by comparing time and spatial scales of the evolution of processes, or more precisely the correlations between variations in time and variations in space, but the issue of the existence of databases precise enough in time appears to be problematic. The study of a link between the derivative of the correlation as a function of window size and of the derivatives of the processes is also a direction to obtain indirect informations on dynamics from static data.

Finally, the search of classes of processes on which it is possible to directly establish the relation between spatial correlations and temporal correlations, is a possible research direction. It stays out of the scope of this present work, but would open relevant perspectives on co-evolution, since it implies evolution in time and an isolation in space, and therefore a complex relation between spatial and temporal covariances.

\subsection{Modeling urban morphogenesis}

We have then proposed a co-evolution model at the mesoscopic scale, based on a multi-modeling paradigm for the evolution of the network. The model is able to reproduce a certain number of observed situations at the first and second order, capturing thus a static representation of interactions between networks and territories. We are therefore able to produce the emergence of the urban form in a coupled way, suggesting the relevance of this approach.

It also yields different dynamical causality regimes, being however less diverse than  for the simple model studied by \cite{raimbault2017identification}: therefore, a more elaborated structure in terms of processes must be paid in flexibility of interaction between these. This suggests a tension between a ``static performance'' and a ``dynamical performance'' of models.

An open question is to what extent a pure network model with preferential attachment for nodes would reproduce results close to what we obtained. The complex coupling between aggregation and diffusion (shown by \cite{2017arXiv170806743R}) could not be easily included, and the model could in any case not answer to questions on the coupling of the dynamics.

\subsection{Implications for policies}

Although remaining theoretical and stylized in the case of the model, our results could give useful insights for policies. First, in terms of empirical computation of correlations, a direct interpretation of their values can inform on the underlying dynamics of the territorial system and suggest issues on which planners must be alerted: for example, a high positive correlation between population hierarchy and network betweenness hierarchy will imply structural congestion effects. The coupling with sustainability indicators such as \cite{le2010approche} does with energy consumption, could also provide useful insights for a sustainable territorial planning. Secondly, a more refined calibration of the model should allow understanding better territorial dynamics in precise cases: for example, focusing on the best network heuristic would provide the most probable processes ruling network evolution, and make experiments on these (e.g. implementing them differently by changing their parameters, or replacing with other processes), given performance indicators. This last point is of a particular interest, as it could provide insights into the combination of top-down (planned) and bottom-up network evolution processes, as both are included in the model.

\section*{Conclusion}

We have introduced a novel methodology to quantify urban form by coupling the built environment (captured by the spatial distribution of population) with road networks, which uses corresponding indicators and their spatial correlations. We studied their empirical values for Europe and the behavior of correlations as a function of scale. This coupled approach was complemented by a morphogenesis model based on a co-evolution paradigm which precisely considers strong relations between both aspects. The calibration of the model on empirical data suggests that such coupled processes indeed occur in the emergence of the urban form and function. The two complementary axis of this work therefore pave the way towards more integrative approaches to urban morphology.

\section*{Acknowledgements}

Results obtained in this paper were computed on the vo.complex-system.eu virtual organization of the European Grid Infrastructure ( http://www.egi.eu ). We thank the European Grid Infrastructure and its supporting National Grid Initiatives (France-Grilles in particular) for providing the technical support and infrastructure.


\end{document}